\documentclass[12pt,preprint]{iau}
\usepackage{graphicx}
\usepackage{color}
\newcommand{\apj}{Astrophys. J. }

\newcommand{\apjs}{Astrophysical Journal Supplement }
\newcommand{\apss}{Astrophysics and Space Science }

\newcommand{\mnras}{Mon. Not. R. Astron. Soc. }
\newcommand{\nar}{New Astronomy Reviews }
\newcommand{\nat}{Nature }
\def\kms{km s$^{-1}$}

\title[Periodic variability of AGN ] %% give here short title %%
{Periodic optical variability of AGN }

\author[Bon et al.]   %% give here short author list %%
{E. Bon$^1$,
 P. Marziani$^2$, N. Bon$^1$
 }
\affiliation{$^1$ Belgrade Observatory, Serbia\\[\affilskip] $^2$INAF, Osservatorio Astronomico di Padova, Italy \\ 
email: {\tt ebon@aob.rs} \\[\affilskip]
}
\pubyear{2016}
\volume{xxx}  %% insert here IAU Symposium No.
\jname{New Frontiers in Black Hole Astrophysics}
\editors{ eds.}
\begin{document}

\maketitle

\begin{abstract}
Here we present the evidence for periodicity of an optical emission detected in several AGN.
Significant periodicity is found in light curves and radial velocity curves.
We discuss possible mechanisms that could produce such periodic variability and their implications.
The results are consistent with possible detection of the orbital motion in proximity of the AGN central supermassive black holes.
\end{abstract}

\firstsection % if your document starts with a section,
\section{Introduction}

Different mechanisms have been proposed to explain active galactic nuclei (AGN) 
emission variability (\cite[see e.g. Gaskell 1983, 2009 and the references 
therein]{Gaskell83,Gask09}).
Besides outflows, jet precession, disk precession, disk warping, spiral arms, 
flares, and other instabilities, one of the most interesting possibilities 
involves the existence of a binary system in the core 
%\cite[Anders \& Zinner (1993)]{AndersZinner93}
(\cite[see eg., Komossa 2006; Bogdanovi\'c et al. 2008; Gaskell
2009; 
%Eracleous et al. 2012; Popovi\'c 2012; 
Bon et al. 2012; Bogdanovi\'c 2015; Graham et al. 2015a, 2015b; Bon et al. 16, Charsi et al. 2016, Liu et al. 2016.  and the references therein)]
{Gaskell83, Komossa, Bog08, Gask09, 
Bon12, Bog15, grahametal15, Graham15b,bonetal16,Charsi2016,Liu2016} 
and the tidal disruption event \cite[(see for e.g. Komossa and Zensus 2015)]{KomossaZensus2015}. 
If black hole (BH) mass grows via major mergers we might expect to see the
signature of a binary black hole in some or many active galaxies, if the merger
process involves slow coalescence of the two components. 

The variability in the nuclei of galaxies was recognized even before AGN were defined as a class. 
There are light curves showing  variability {in} over 100 years of observations in some  objects, for example NGC4151 \cite[(from 1906, see e.g., Oknyanskij 2007)]{Oknyanskij2007}),
3c273 (\cite[from 1880s Smith \& Hoffleit 1963]{SmitHoff1963}), 
with long variability timescale of more than one decade. In the 1950s it was 
reported that the NGC 5548 nucleus  appeared to vary by about 
a magnitude \cite[(see more in Gaskell \& Klimek 2003)]{Gask03}.

It is well known that many AGN show variability in different 
time scales and all wavelengths \cite[(Gaskell \& Klimek 2003)]{Gask03}.
The variability timescales are affected by 
speed at which physical processes propagate; most relevant are
 the speed of light c $\sim$ 3 $\cdot$ 10$^5$ \kms, the orbital 
speed $v_{orb}\sim(GM/R)^{1/2}$ and the
sound speed $v_s\sim(kT/m)^{1/2}$, where $c > v_{orb} >> v_s$.  
The shortest timescale corresponds to the light crossing timescale. 
Reverberation campaigns  are based on measuring  light travel time between 
 a spectral feature representative of the ionizing continuum and an emission line. If the time delays are measured as a function of radial velocity, it is possible at least in principle  to map the line emitting region.   

Orbital timescales are longer \cite[(Netzer 2013)]{Netzer2013}. Measuring them could be  very useful for determining the dynamical mass of the central BH,  while sound speed timescale can give us  information about the property of the disk or the broad line region (BLR).

A question arises: can we identify periodic variations? Such periodic variability should correspond to orbital motion exclusively,  while the other two processes could produce only quasi periodic signals.
Having an estimate of the BH masses, we can easily filter  expected domains of periodicity that could 
correspond to the orbital motion, for the optical spectral  range, where we have the longest observing records.  Unfortunately, the brightest  active galactic nuclei were identified only about 
70 years ago. Therefore observing records are 
relatively short, and do not allow to trace many orbits in historical light curves. 
Extensive spectroscopic observations 
started even later, because the brightest AGN were still faint for the instrumentation available till the mid 1970s. With such limitation in the data, it is really hard to prove that the signal 
is  periodic. If we were able to reveal periodic signals on the orbital 
timescales, combining results from 
the light crossing timescales, and measuring 
the broad emission line shapes 
(since we know that the low-ionization part of the emitting region is virialized)  we could   really be able to map the BLR.

\section{Light curve variability patterns}
\label{rec}

AGN variability patterns present in their light curves mainly correspond to  red noise, 
giving spuriously high significance levels to low frequency periods 
(see Westman et al. 2011, Vaughan et al. 2016 and Bon et al. 2016). 

AGN have been since long known to be dominated by red noise-like light curves. 
Therefore, standard methods, such as the Lomb-Scargle (LS) method \cite[(Lomb 1976, Scargle 1982)]{lomb76, scarg82} 
may show very high peaks but the  probability of a false positive computed with conventional statistical 
approximations  may not be valid.   Such problems in using LS periodogram tests for the analysis of AGN X-ray 
light curves  has been known   for some time, and they appear especially serious for very frequent quasi-periodic 
variability detections over long time-scales which turned out to be false 
(e.g., see discussion in Vangham 2006).

AGN light curves can be modeled  as damped random walks (DRW) or with autoregressive models  
 \cite[(as it has been known since many years; e.g., Fahlman \& Ulrych 1975, Gaskell \& Peterson 1987)]
 {Fahlman1975, GasPet1987}. There are some attempts to systematically search for periods in quasars LCs 
 taking this problem into account \cite[(e.g. see Charsi et al 2016, Liu et al. 2016)]{Charsi2016, Liu2016}. 
These studies show that periodicities of some candidates show very significant power-spectrum peaks compared to 
expectation for red noise.  In some cases there is evidence of even more then one period with very high significance  
compared to autoregressive (AR) expectations \cite[(see e.g. Bahta et al. 2016)]{Bahta}.

\begin{figure}[b]
% \vspace*{-0.5 cm}
\begin{center}
 \includegraphics[width=5in]{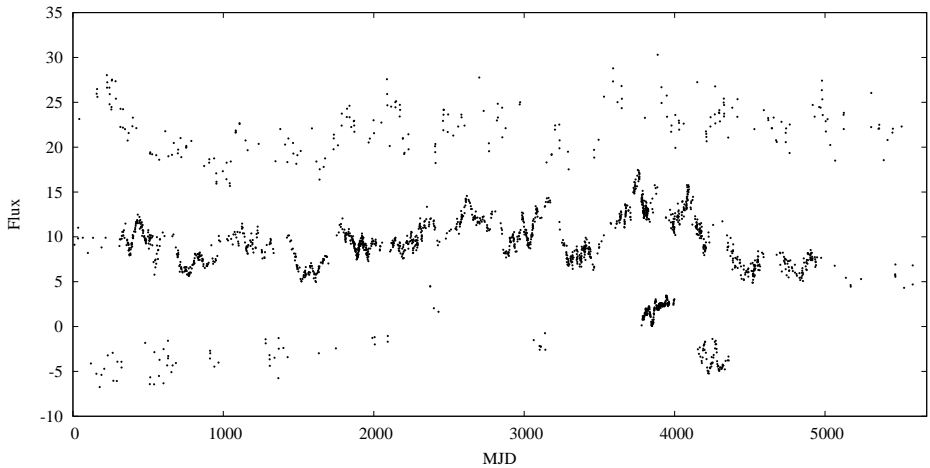} 
% \vspace{0.5 cm}
 \caption{Folded {phased} diagram of NGC 5548 continuum flux 45 yr long light curve, {assuming} 5676 days periodicity, 
 here presented to show an example of repeating noisy pattern light curve. {Top light curve is pre AGN watch epoch, 
 middle is AGN watch epoch \cite{Sergeev2007}, and bottom is post AGN watch epoch constructed from new 
 observations presented in \cite{bonetal16, Lu2016, Pei2017}.}}
   \label{fig1}
\end{center}
\end{figure}

\begin{figure}[b]
% \vspace*{-0.5 cm}
\begin{center}
 \includegraphics[width=3.5in,angle=270]{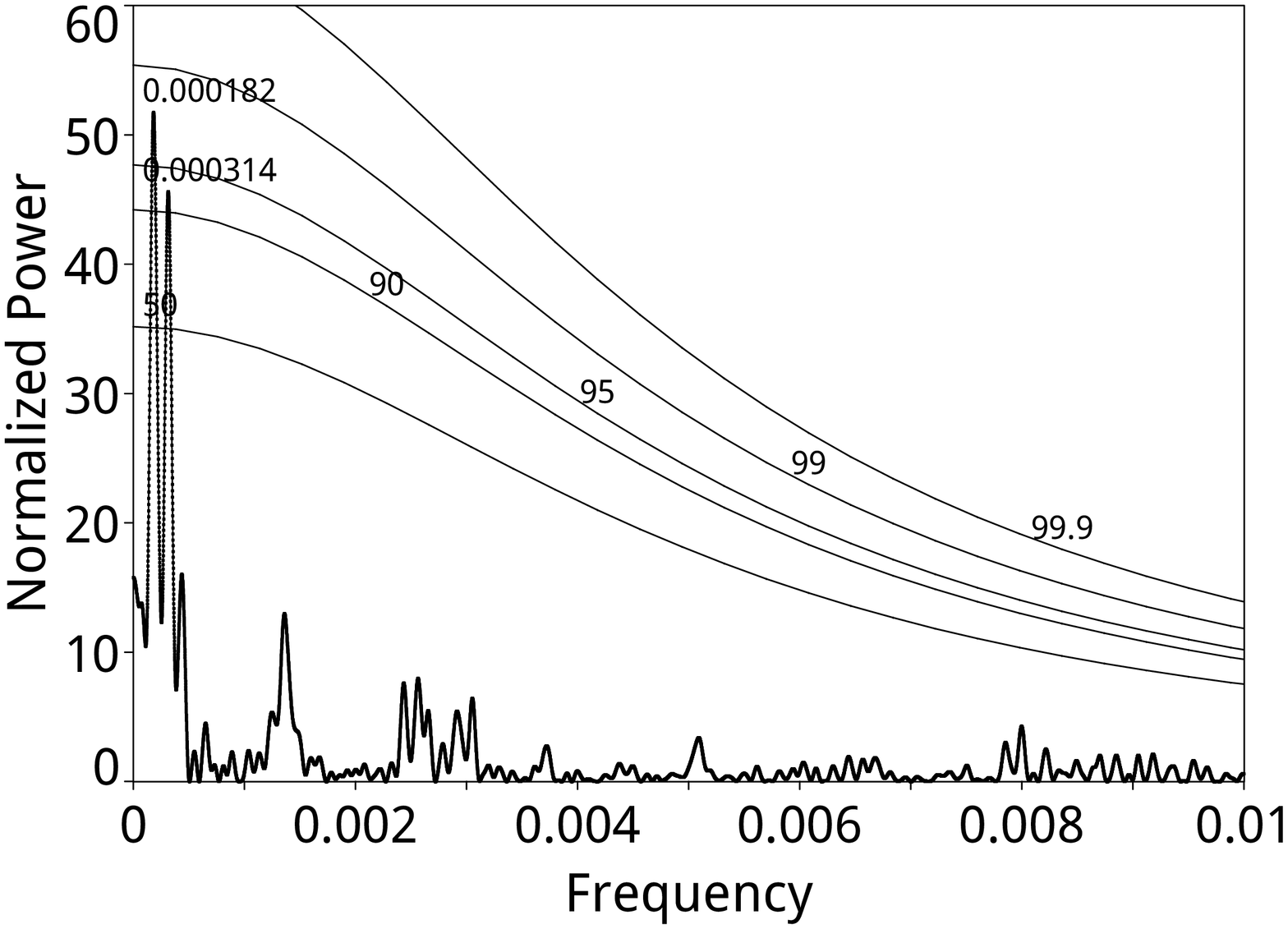}
 % \vspace*{-1.0 cm}
 \caption{Example of Lomb-Scargle periodogram of NGC 5548 radial velocity curve constructed 
from the red side broad H$\beta$ half-widths measured at 75\% of maximum \cite[(see Bon et al. 2016)]{bonetal16}. 
Red noise {background spectrum} significance levels are marked.}
   \label{fig2}
\end{center}
\end{figure}

Nevertheless, even though activity within one period might look stochastic compared to the DRW tests, 
repeating patterns could appear on very long light curves, showing the same complex patterns   
starting to repeat themselves after some periodic interval (see folded light curve 
presented in Fig.\,\ref{fig1}, where one can see that noisy pattern 
repeated after 5676 days in the NGC 5548 optical light curve; more details are 
available in \cite[Bon et al 2016]{bonetal16}). 
Periodicity was found using a new method,  similar to phase dispersion minimization  
\cite[({PDM, see} Stellingwerf 1978)]{Stellingwerf1978},
specially developed for the uneven-sampled 43 yr 
data series  of NGC 5548 \cite[(Bon et al. 2016)]{bonetal16}. 
Beside  light curves, radial velocity curves of NGC 5548 measured on the broad H$\beta$  profile %also 
show periodicity  
%that {\bf is} significant 
{of 5550 d, with a peak above 95\% 
AR significance level (of LS false alarm probability)},
%if compared to red noise expectations, 
as shown in  Fig.\,\ref{fig2}, 
(see Table 3 in \cite[Bon et al. 2016]{bonetal16}). 
{Using a PDM method and the new compiled light curve from Fig.\,\ref{fig1} we found significant peak at 5678 d}.

%\section{Radial velocity curves}

Until now there are only few candidates with the detection of the same periodicity in both  light and radial velocity curves:  
e. g., NGC 4151 (Bon et al. 2012) and NGC 5548 (Li et al 2016, Bon et al. 2016).  
Both candidates have been monitored for very long time intervals, with probably the longest and best sampled data series of 
spectra and light curves, spanning over 43 years for NGC5548 and over 25 yr for NGC4151. 

\section{Physical interpretations}

Probably the most plausible scenario for relatively  short periodicity is  orbital motion
\cite[(see eg. Bon et al. 2012, 2016; Graham et al. 2015a, 2015b{)}]{bonetal12, bonetal16, grahametal15, Graham15b}. 
{Some mechanisms, 
%as discussed in \cite[Bon et al. 2012, 2016]{\bonetal12, bonetal16}, 
are geodetic precession 
\cite[(Begelman et al. 1980)]{Begelman80}, 
accretion disk precession} due to a second black hole \cite[(Katz 1997)]{Katz97}, and disk-self warping induced by 
radiation pressure \cite[(Pringle 1996)]{Pringle96}.
Some of these periodic AGN are recognized as SMBBH candidates, like NGC4151 \cite[(Bon et al. 2012)]{bonetal12}, 
OJ287 \cite[(e.g. Pihajoki 2016)]{Pihajoki2016, Bahta}, PG1302-102 \cite[(Graham et al. 2015a)]{grahametal15}.  
Geodetic precession is excluded since it occurs on time scales that are
in general much longer than currently available monitoring times. 
%One possibility is that the orbital 
Periodicity could be  due to the  presence of a long-lived hotspot in the disk \cite[(Jovanovi\'c  et al. 2010)]{Jovanovic10}. 
We note however  that  the flare lasted only
about a year, and was spiraling outwards in the model of \cite[Jovanovi\'c  et al. (2010)]{Jovanovic10}. 
Therefore, it would be more likely that the hotspot could be periodically generated by an orbiting body 
passing through the accretion disk \cite[(eg. 
Kieffer \& Bogdanovic 2016, Pihajoki 2016, Bon et al. 2016)]{Bogdanovic2016, Pihajoki2016, bonetal16}, 
with an orbital period similar to S0 stars in our Galaxy, or alternatively, a periodic enhancement 
of the inner part of the disk producing the X-ray emission 
%\cite[(Fausnaugh et al 2015, Edelson et al. 2015, Bon et al. 2016)]{{Fausnaugh2015, Edelson2015,bonetal16}}.
%\cite[(Bon et al. 2016)]{{bonetal16}}.
\cite[(Fausnaugh et al 2015, Bon et al. 2016)]{{Fausnaugh2015, bonetal16}}.

\cite[Graham et al. 2015b]{Graham et al. (2015a)} calculated that warped disks around single black holes are not favored in an
AGN context unless very special conditions are satisfied. If the accretion disk is warped, it could
lead to precession of the jet, but again with periodicity much longer then current monitored time. 
Yet another alternative possibility could be an obscuration of the
central accretion disk, and of part of the BLR by a moving obscuring object \cite[(Bon et al. 2016)]{bonetal16},
%Gaskell intrinsic reddening
{or moving emitting region, since BLR could consist of two different emitting components
%: disk like region, and vertical component 
%that might affect the core of the broad line 
\cite[(Bon 2008, Bon et al. 2009a,2009b, 2015)]{Bon08, Bon09a, Bon09b, Bon2015}.}
%or the blue side of the line \cite{Bon2015}.
%\vspace*{-.25 cm}

%\bibliographystyle{iau}
%\bibliography{biblioletter2}

\end{document}